\documentclass[12pt]{article}
\usepackage{graphicx}
\usepackage{amsmath}
\usepackage{amssymb}

\numberwithin{equation}{section}

\begin{document}
\setlength{\baselineskip}{18pt}
\begin{titlepage}
\begin{flushright}
{\small OU-HET 604/2008}\\[-1mm]
{\small KUNS-2137}%
\end{flushright}

\vspace*{5mm}
\begin{center}
{\Large Tri-bimaximal Mixing from Cascades}
\end{center}
\vspace*{5mm}

\begin{center}
Naoyuki Haba$^1$, Ryo Takahashi$^1$, Morimitsu Tanimoto$^2$, and
Koichi Yoshioka$^3$
\vspace{7mm}

$^1${\it Department of Physics, Graduate School of Science, 
Osaka University,\\[-1mm]
Toyonaka, Osaka 560-0043, Japan}\\
$^2${\it Department of Physics, Niigata University, 
Niigata 950-2181, Japan}\\
$^3${\it Department of Physics, Kyoto University,
Kyoto 606-8502, Japan}
\end{center}

\vspace*{12mm}
\begin{abstract}\noindent
We study fermion mass matrices of the cascade form which are
compatible with the tri-bimaximal lepton mixing and generation mass
hierarchy. The flat-cascade lepton matrices imply a 
parameter-independent relation among the mixing angles and mass
eigenvalues. The relation has several indications that the atmospheric
neutrino mixing angle is close to maximal and the other two angles
have a correlation independently of neutrino mass eigenvalues. We also
discuss phenomenological aspects of the cascade matrices;
flavor-violating rare decays of charged leptons, thermal leptogenesis,
and leptonic CP violation. Possible dynamical origins of the cascades
are illustrated based on flavor symmetry and in higher-dimensional
theory.
\end{abstract}

\end{titlepage}

\newpage
\section{Introduction} 

Neutrino physics is one of the most important clues to seek further
physics beyond the standard model (SM)\@. The neutrino oscillation
experiments are going into a new phase of precision measurements of
generation mixing angles and mass squared differences. The generation
mixing in the lepton sector has been found to be quite different from
that in the quark sector: there are large mixings among the
three-generation leptons. Various recent observations have been
indicating that the experimental data of lepton mixing converses to
the tri-bimaximal form~\cite{TBM,TBM2}, which is given by
\begin{eqnarray}
  V_{\rm TB} \;=\; \left(\begin{array}{ccc}
  \frac{2}{\sqrt{6}}  & \frac{1}{\sqrt{3}} & 0 \\[1mm]
  \frac{-1}{\sqrt{6}} & \frac{1}{\sqrt{3}} & \frac{-1}{\sqrt{2}} \\[1mm]
  \frac{-1}{\sqrt{6}} & \frac{1}{\sqrt{3}} & \frac{1}{\sqrt{2}}
  \end{array}\right),
\end{eqnarray}
up to complex phases of light neutrino mass eigenvalues. The current
experimental data~\cite{EXP} of mixing angles is well approximated 
by $V_{\rm TB}$ and in turn implies a specific form of mass matrix for
light neutrinos. For light Majorana-type neutrinos, the mass matrix in
the flavor basis of $e$, $\mu$ and $\tau$ becomes
\begin{eqnarray}
M_L &=& V_{\rm TB}^*
\left(\begin{array}{ccc}
  m_1 & & \\
  & m_2 & \\
  & & m_3
\end{array}\right)
V_{\rm TB}^\dagger  \nonumber \\
&=& \frac{m_1}{6}\left(\begin{array}{ccc}
  4  & -2 & -2 \\
  -2 & 1  & 1  \\
  -2 & 1  & 1
\end{array}\right)+
\frac{m_2}{3}\left(\begin{array}{ccc}
  1 & 1 & 1 \\
  1 & 1 & 1 \\
  1 & 1 & 1 
\end{array}\right)+
\frac{m_3}{2}\left(\begin{array}{ccc}
  0 & 0  & 0  \\
  0 & 1  & -1 \\
  0 & -1 & 1
\end{array}\right),
\label{MLTB}
\end{eqnarray}
where $m_{1,2,3}$ are the mass eigenvalues of light neutrinos. It is
found from this expression that the experimentally favored neutrino
matrix is restricted to a special form in which the matrix elements
are integer (inter-family related) valued. Such a suggestive form
seems to indicate a hidden structure in nature beyond the SM, and a
number of proposals to unravel it have been 
elaborated~\cite{TBMmodels}.

In this paper, we investigate the neutrino and charged-lepton mass
matrices in the cascade form. While the cascade-form matrix has
hierarchical orders of matrix elements in generation space, it
can generate large lepton mixing, in particular, the
tri-bimaximal mixing in the lepton sector, as will be shown in
Section~\ref{cas_lep}. Such a compatibility of large generation
mixing with mass hierarchy is suitable for the extensions to the quark
sector and grand unification. The cascade-form matrix implies a
parameter-independent relation among the lepton mixing angles and
mass eigenvalues, which would be tested in future neutrino
oscillation experiments. In Section~\ref{pheno}, several
phenomenological aspects of the cascade lepton matrices are also
discussed, e.g.\ the lepton flavor-violating processes, the thermal
leptogenesis, and the CP violation in neutrino oscillations. In
Section~\ref{dynamics} we present possible dynamical origins of
cascades in flavor symmetric theory and in higher-dimensional
spacetime. Section~\ref{summary} is devoted to summarizing the results.

\bigskip

\section{Cascade matrix}

\begin{figure}[t]
\begin{center}
\begin{minipage}{4.5cm}
\includegraphics[width=4.3cm,height=4.8cm]{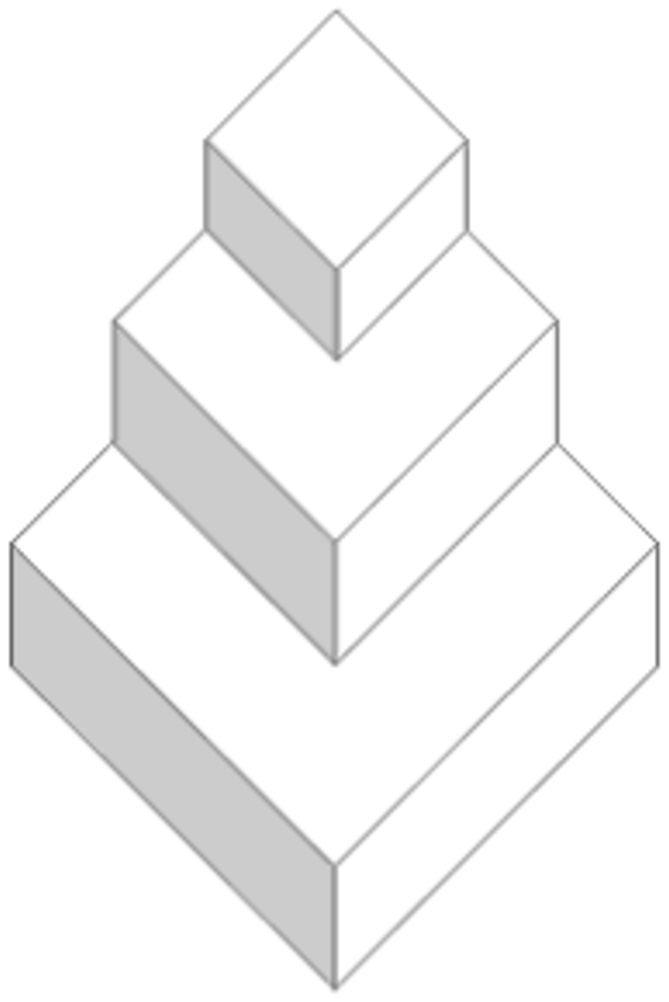}
\vspace*{10mm}

\centering{[$\,$Cascade$\,$]\bigskip}
\end{minipage}
\hspace*{3mm}
\begin{minipage}{4.5cm}
\includegraphics[width=4.5cm,height=5.8cm]{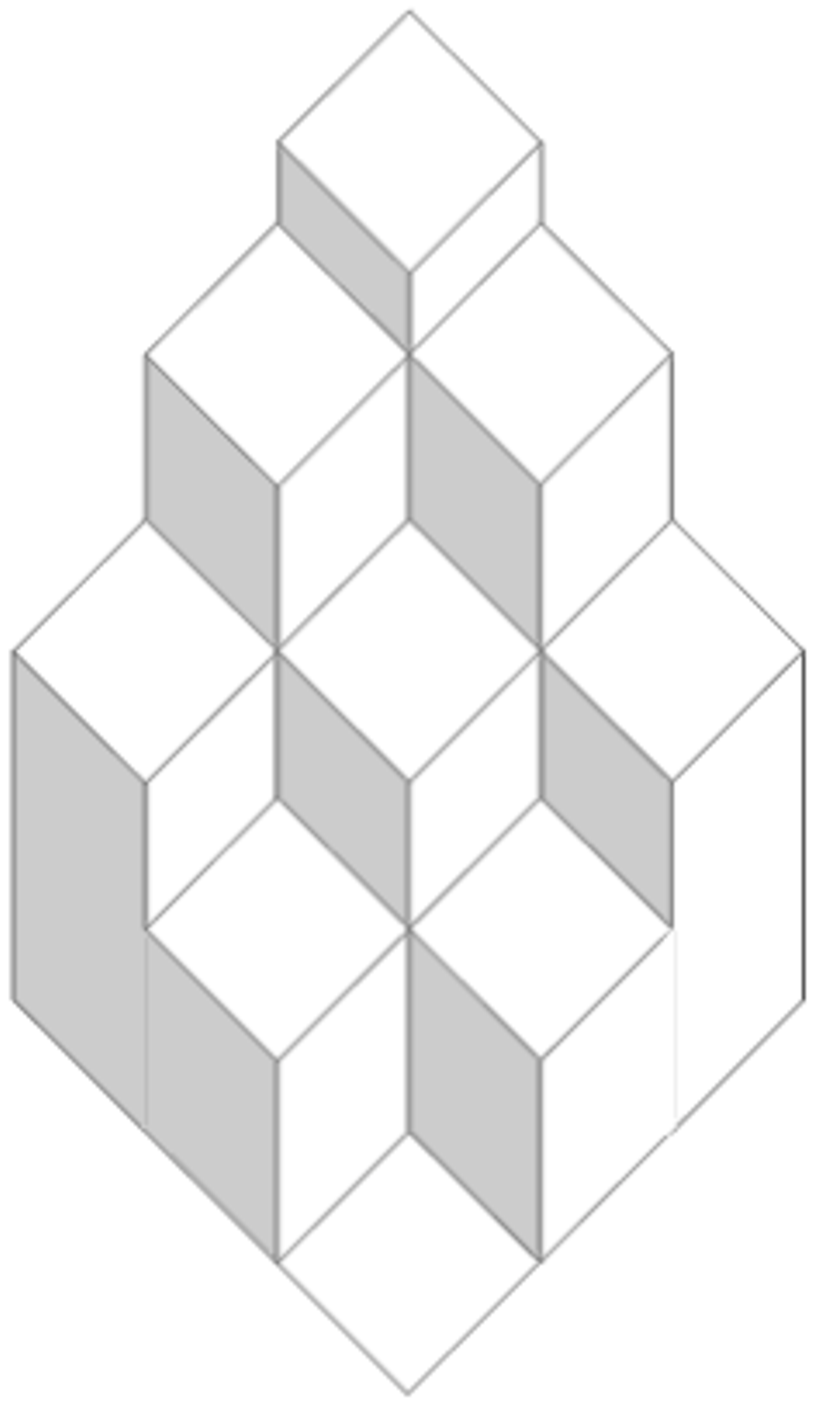}
\centering{[$\,$Waterfall$\,$]\bigskip}
\end{minipage}
\hspace*{4mm}
\begin{minipage}{4.5cm}
\includegraphics[width=4.5cm,height=5cm]{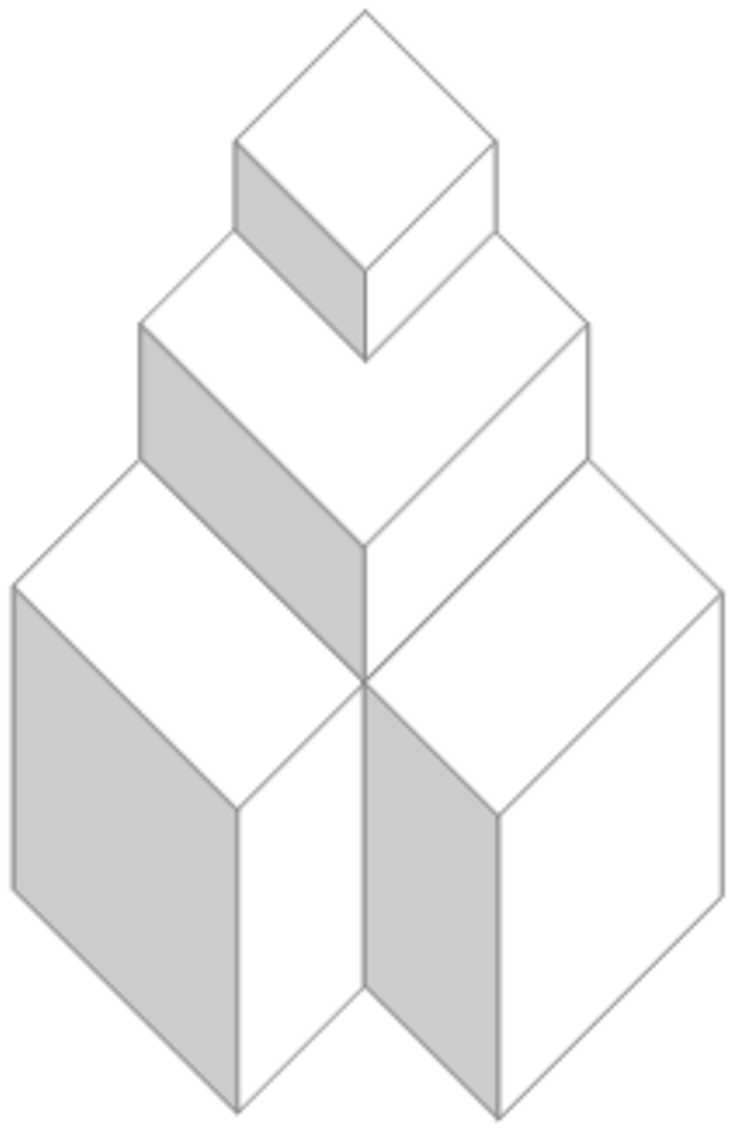}
\vspace*{3mm}

\centering{[$\,$Hybrid$\,$]\bigskip}
\end{minipage}
\caption{Hierarchy flows of matrix elements.\bigskip}
\label{fig:matrices}
\end{center}
\end{figure}

In this paper we investigate the following form of mass matrix:
\begin{eqnarray}
  M_{\rm cas} \;=\; \left(\begin{array}{ccc}
   \delta & \delta & \delta \\
   \delta & \lambda & \lambda \\
   \delta & \lambda & 1 
  \end{array}\right)v
  \label{Mcas}
\end{eqnarray}
with the small parameters $|\delta|\ll|\lambda|\ll1$. The 
dimension-one parameter $v$ denotes the overall mass scale and is
given by some scalar expectation value times the largest element of
Yukawa matrix. There are generally ${\cal O}(1)$ coefficients
in the matrix elements, not explicitly written in the above, and 
so $M_{\rm cas}$ is not necessarily left-right symmetric. The 
matrix \eqref{Mcas} is called the cascade form in the view of its
hierarchical structure of matrix elements (see 
Fig.~\ref{fig:matrices}). To clarify the property of cascade matrix,
we will show in parallel the results of the following matrix form
which has been well studied in the literature:
\begin{eqnarray}
  M_{\rm wat} \;=\; \left(\begin{array}{ccc}
    \delta^2 & \delta\lambda & \delta \\
    \delta\lambda & \lambda^2 & \lambda \\
    \delta & \lambda & 1 
  \end{array}\right)v,
  \label{Mwat}
\end{eqnarray} 
where ${\cal O}(1)$ coefficients have also been dropped in the matrix
elements. For comparison, the generation mixings are set to be of the
same order between the above two types of matrices. The mass 
matrix \eqref{Mwat} has a more rapid stream of hierarchy flow than the
cascade one (see Fig.~\ref{fig:matrices}) and is called here the
waterfall mass matrix.
The waterfall matrix is realized, for example, in the Froggatt-Nielsen
model~\cite{FN} with abelian flavor symmetry, where the mass terms are
effectively induced from higher-dimensional operators including a
scalar field $\phi$ whose expectation value is smaller than some cutoff 
scale $\Lambda$; $\,\langle\phi\rangle/\Lambda\equiv\rho\,\ll1$. Given
a quantum charge assignment $Q_{1,2,3}$ for the three-generation
fields, the matrix \eqref{Mwat} is interpreted 
as $\delta\sim\rho^{Q_1}$, $\,\lambda\sim\rho^{Q_2}$, 
and $1\sim\rho^{Q_3}$. Here we have taken the equal charges for left
and right-handed fermions, for simplicity. The examples of dynamics
for $M_{\rm cas}$ will be discussed in a later section by use of flavor
symmetry.

The two types of matrices have the same orders of generation mixing
angles $\theta_{12}$, $\theta_{23}$, $\theta_{13}$, while they induce
different mass eigenvalues $m_{1,2,3}$, as shown in the following
table:\smallskip
\begin{eqnarray}
{\renewcommand{\arraystretch}{1.3}%
\quad\begin{array}{c|c|c} \hline
& M_{\rm cas} & M_{\rm wat} \\ \hline\hline
\text{masses}~ 
& m_1:m_2:m_3 \;\sim\; \delta : \lambda : 1
& m_1:m_2:m_3 \;\sim\; \delta^2 : \lambda^2 : 1 \\ \hline
\text{mixing}~ & 
~\theta_{12}\sim \delta/\lambda, \;\;\;
\theta_{23}\sim \lambda, \;\;\;
\theta_{13}\sim \delta~ & 
~\theta_{12}\sim \delta/\lambda, \;\;\;
\theta_{23}\sim \lambda, \;\;\;
\theta_{13}\sim \delta~ \\[1mm]
& \Big(\;\theta_{ij}\sim \frac{m_i}{m_j}\;\Big) 
& \Big(\;\theta_{ij}\sim \sqrt{\frac{m_i}{m_j}}\;\Big) \\[2mm] \hline
\end{array}}
\label{McasMwat}
\end{eqnarray}
The relations between the eigenvalues and mixing angles are given 
by $\theta_{ij}\sim m_i/m_j$ for the cascade matrix 
and $\theta_{ij}\sim\sqrt{m_i/m_j}$ for the waterfall matrix. It is
interesting here to remember the well-known relations among the quark
generation mixing and the down-type quark masses $m_{d_i}$. The
experimentally observed values of quark masses and mixing angles are
roughly related as
\begin{equation}
  \theta_{12}^q \,\sim\, \sqrt{\frac{m_{d_1}}{m_{d_2}}}, \qquad
  \theta_{23}^q \,\sim\, \frac{m_{d_2}}{m_{d_3}}, \qquad
  \theta_{13}^q \,\sim\, \theta_{12}^q\theta_{23}^q.
\end{equation}
It is found from these expressions that $\theta_{12}^q$ is induced
from a waterfall matrix, while $\theta_{23}^q$ is described by a
cascade matrix. This hybrid pattern can be achieved if the cascade
form is slightly modified, that is, the 1-1 matrix element is made
vanishing. The hybrid cascade matrix takes the form
\begin{eqnarray}
  M_{\rm hyb} \;=\; \left(\begin{array}{ccc}
   0 & \delta & \delta \\
   \delta & \lambda & \lambda \\
   \delta & \lambda & 1 
  \end{array}\right)v.
\end{eqnarray}
The generation mixing is set to be of the same order of the previous
two matrices $M_{\rm cas}$ and $M_{\rm wat}$. The resultant mass
hierarchy and mixing angles are estimated as\smallskip
\begin{eqnarray}
{\renewcommand{\arraystretch}{1.3}%
\qquad\begin{array}{c|c}\hline
  & M_{\rm hyb} \\ \hline\hline
\text{masses}~ & 
m_1:m_2:m_3 \;\sim\; \delta^2/\lambda : \lambda : 1 \\ \hline
\text{mixing}~ & 
\theta_{12}\sim \delta/\lambda, \;\;\;
\theta_{23}\sim \lambda, \;\;\;
\theta_{13}\sim \delta  \\[1mm]
& ~\Big(\,\theta_{12}\sim \sqrt{\frac{m_1}{m_2}}, \;\;
\theta_{23}\sim \frac{m_2}{m_3}, \;\;
\theta_{13}\sim \frac{\sqrt{m_1m_2}}{m_3}\,\Big)~ \\[2mm] \hline
\end{array}}
\end{eqnarray}
That implies the (modified) cascade form is viable in the quark
sector. Furthermore the cascade matrix is suitable for the lepton
sector, as we will discuss in the following sections. Thus the cascade
form matrix is expected to be embedded into grand unified theory.

\bigskip

\section{Cascade lepton matrices}
\label{cas_lep}
\subsection{Neutrino sector}

We first consider the situation that the neutrino Dirac mass matrix
takes the cascade form:
\begin{eqnarray}
  M_N \;=\; \left(\begin{array}{ccc}
    \delta_1 & \delta_2 & \delta_3 \\
    \delta_2 & \lambda_1 & \lambda_2 \\
    \delta_3 & \lambda_2 & 1
  \end{array}\right)v
 \label{cas}
\end{eqnarray}
with the mass parameter 
hierarchy $|\delta_i|\ll|\lambda_j|\ll1$, having in mind the extension
to more fundamental theory including quarks and grand
unification. Throughout this paper the cascade matrix is assumed to be
left-right symmetric, which is the simplest example and may be
preferable to be realized. For the Majorana mass matrix of
right-handed neutrinos, we have
\begin{eqnarray}
  M_R \;=\; \left(\begin{array}{ccc}
    \!M_1 & & \\
    & \!M_2\! & \\
    & & M_3\!
  \end{array}\right).
\end{eqnarray}
It is noted that, if one would assume that $M_R$ is also of the
cascade form, the following results in this paper do not change 
qualitatively. This is because, as we will show, the right-handed
neutrino masses are experimentally required to have larger hierarchy
than the Dirac masses and therefore the generation mixing (the
off-diagonal elements) in $M_R$ becomes negligible. Accordingly we
are allowed to take the diagonal form of $M_R$ from the beginning.

The first task is to find experimental indications on the cascade
neutrino matrices, referring to the current experimental data of
neutrino mass eigenvalues and mixing. (The charged-lepton contribution
to the lepton generation mixing will be included in the next 
section.) \ After integrating out the heavy right-handed 
neutrinos~\cite{seesaw}, one obtains the Majorana mass matrix for
three generations of light neutrinos in low-energy effective theory:
\begin{equation}
  M_L \;=\;
  \frac{v^2}{M_1}\!\left(\begin{array}{ccc}
    \!\delta_1^2 & \delta_1\delta_2 & \delta_1\delta_3\! \\
    \!\delta_1\delta_2 & \delta_2^2 & \delta_2\delta_3\! \\
    \!\delta_1\delta_3 & \delta_2\delta_3 & \delta_3^2\! 
  \end{array}\right)+
  \frac{v^2}{M_2}\!\left(\begin{array}{ccc}
    \!\delta_2^2 & \delta_2\lambda_1 & \delta_2\lambda_2\! \\
    \!\delta_2\lambda_1 & \lambda_1^2 & \lambda_1\lambda_2\! \\
    \!\delta_2\lambda_2 & \lambda_1\lambda_2 & \lambda_2^2\!
  \end{array}\right)+
  \frac{v^2}{M_3}\!\left(\begin{array}{ccc}
    \!\delta_3^2 & \delta_3\lambda_2 & \delta_3\! \\
    \!\delta_3\lambda_2 & \lambda_2^2 & \lambda_2\! \\
    \!\delta_3 & \lambda_2 & 1\!
  \end{array}\right).\;
  \label{ML}
\end{equation}
Comparing this with the experimentally favored form \eqref{MLTB} and
taking into account the cascade 
hierarchy $|\delta_i|\ll|\lambda_j|\ll1$, we are approximately lead to
the following relations among the parameters:
\begin{eqnarray}
  \delta_1\,=\,\delta_2\,=\,\delta_3\;\;\,(\equiv\delta)\,, \qquad
  \lambda_1\,=-\lambda_2\;\;\,(\equiv\lambda)\,.
  \label{cond}
\end{eqnarray}
These are not the claims of fine tuning but should be interpreted as
a first approximation for the current experimental data (remember
that the tri-bimaximal generation mixing is almost at the center of
the experimentally allowed region of parameter space). Such types of
parameter relations have often been seen in the lepton mass 
models, e.g.\ with the vacuum alignments and non-abelian flavor
symmetry which connects different generations~\cite{LMmodels}. In this
paper, we study phenomenological results of the cascade lepton
matrices with \eqref{cond} as the first approximation in a suggestive
form, and discuss its characteristic property in current and future
particle/cosmological experiments such as for lepton flavor and CP
violations. Later (in Section~\ref{dynamics}), we will present several
flavor symmetry dynamics for the cascade-form matrix.

The mass eigenvalues are roughly given by 
${m_\nu}_1\sim v^2/M_3$, ${m_\nu}_2\sim\delta^2 v^2/M_1$, and 
${m_\nu}_3\sim\lambda^2v^2/M_2$. These masses (and the tri-bimaximal
generation mixing) are perturbed by the small 
quantities ${m_\nu}_1/{m_\nu}_{2,3}$ and $\delta/\lambda$. The
cascade neutrino model has the normal hierarchy of light neutrino mass
spectrum, and the mass eigenvalues are explicitly given by
\begin{eqnarray}
  {m_\nu}_1 &=& \frac{v^2}{6M_3}, \\
  {m_\nu}_2 &=& \frac{v^2}{3M_3}+\frac{3\delta^2v^2}{M_1}, \\
  {m_\nu}_3 &=& \frac{v^2}{2M_3}+\frac{2\lambda^2v^2}{M_2},
\end{eqnarray}
including the leading-order corrections of ${\cal O}({m_\nu}_1)$. On
the same order of perturbation evaluation, the effective neutrino mass
matrix is almost diagonalized by the tri-bimaximal mixing
matrix. Small deviations are evaluated at the first order in
perturbation theory and the mixing angles are determined as follows:
\begin{eqnarray}
  \sin^2\theta_{12} &=& \bigg|\frac{1}{\sqrt{3}}
  -\frac{2}{\sqrt{3}}\frac{{m_\nu}_1}{{m_\nu}_2}\bigg|^2, \\[1mm]
  \sin^2\theta_{23} &=& \bigg|\frac{-1}{\sqrt{2}} 
  +\frac{1}{\sqrt{2}}\frac{{m_\nu}_1(3{m_\nu}_3
    -{m_\nu}_2)}{{m_\nu}_3({m_\nu}_3-{m_\nu}_2)} 
  +\frac{\delta}{3\sqrt{2}\lambda}\,
  \frac{{m_\nu}_2}{{m_\nu}_3-{m_\nu}_2}\bigg|^2, \\[1mm]
  \sin^2\theta_{13} &=& \bigg|\frac{\delta}{\sqrt{2}\lambda}\,
  \frac{{m_\nu}_3-\frac{2}{3}{m_\nu}_2}{{m_\nu}_3-{m_\nu}_2}
  +\frac{\sqrt{2}\,{m_\nu}_1{m_\nu}_2}{{m_\nu}_3({m_\nu}_3-{m_\nu}_2)}
  \bigg|^2.
\end{eqnarray}

The hierarchical Dirac neutrino mass matrix of the cascade form
induces the large generation mixing; the first term in \eqref{ML} and
the first relation in \eqref{cond} means the tri-maximal mixing of
three generations, and the second term in \eqref{ML} and the second
equality in \eqref{cond} implies the bi-maximal mixing of the second
and third generations where the cascade 
hierarchy $|\delta|\ll|\lambda|$ plays an important role. These
together uniquely define the unitary mixing matrix of the
tri-bimaximal form. The remaining third term in \eqref{ML} has tiny
generation mixing due to the mass hierarchy and only gives small 
corrections. In the waterfall model widely studied with $U(1)$ flavor
symmetry, the tri-maximal and/or bi-maximal nature seems not to be
simply captured, since the steepness in every point of the stream
generally requires an elaborate form of right-handed neutrino mass
matrix, which might be difficult to rely on some theoretical
background.

As for high-energy couplings before the seesaw, the cascade hierarchy
parameters are loosely bounded as
\begin{eqnarray}
  \bigg|\frac{\delta}{\lambda}\bigg|^2 &\ll&
  \frac{\Delta m_{21}^2}{\Delta m_{31}^2},
  \label{bound}
\end{eqnarray}
in order for the model to be consistent with the observed generation
mixing. Here $\Delta m_{21}^2\equiv|{m_\nu}_2|^2-|{m_\nu}_1|^2$ 
and $\Delta m_{31}^2\equiv|{m_\nu}_3|^2-|{m_\nu}_1|^2$ are the mass
squared differences of light neutrinos and the current experimental
data~\cite{EXP} at the $3$ sigma level is
\begin{alignat}{2}
  \sin^2\theta_{12} &\;=\; 0.32^{\,+0.08}_{\,-0.06}\,,
  & \hspace*{20mm}
  \Delta m_{21}^2\, &\;=\; 7.6^{\,+0.7}_{\,-0.5}
  \times 10^{-5} \;{\rm eV}^2,  \nonumber \\
  \sin^2\theta_{23} &\;=\; 0.50^{\,+0.17}_{\,-0.16}\,,&
  |\Delta m_{31}^2| &\;=\; 2.4^{\,+0.4}_{\,-0.4}
  \times 10^{-3} \;{\rm eV}^2, \nonumber \\
  \sin^2\theta_{13} &\;<\; 0.050\,. & &
\end{alignat}
Then the bound \eqref{bound} means that the ratio between the two
cascade falls is $|\delta|/|\lambda|\ll0.16-0.20$. There is no
experimental upper bound on $|\lambda|$, and the cascade could have a
mild hierarchy.

As for the right-handed neutrino Majorana masses, they are estimated
from the experimental data of neutrino oscillations. Given the normal
hierarchy of light neutrino mass spectrum, the right-handed neutrino
masses become
\begin{eqnarray}
  |M_1| &\simeq& 3.4\times10^{11}
  \,|\delta|^2\left(\frac{v}{\mbox{GeV}}\right)^2\;\mbox{GeV},
  \label{M1} \\
  |M_2| &\simeq& 4.0\times10^{10}
  \,|\lambda|^2\left(\frac{v}{\mbox{GeV}}\right)^2\;\mbox{GeV},
  \label{M2} \\
  |M_3| &\simeq& 1.9\times10^{12}\,
  \left(\frac{v}{\mbox{GeV}}\right)^2\;\mbox{GeV},
  \label{M3}
\end{eqnarray}
with the best fit values of the experimental data (and no complex
phase parameters assumed). The first two generation 
masses, $M_1$ and $M_2$, are determined independently of the
tri-bimaximal generation mixing. The third-generation mass $M_3$ does
not have theoretical and/or experimental upper bound, and the
limit $|M_3|\to\infty$ means that the lightest 
eigenvalue ${m_\nu}_1$ vanishes and the tri-bimaximal generation
mixing is achieved with a large cascade hierarchy. On the other 
hand, $M_3$ has a lower bound which is given by the maximal deviations 
of $\sin^2\theta_{12}$ and $\Delta m_{21}^2$ from their best fit
values, i.e.\ $|M_3|\geq3.8\times10^{11}\,(v/\mbox{GeV})^2$ GeV\@. We
thus find the right-handed neutrinos generally have the mass
hierarchy $|M_1|<|M_2|\ll|M_3|$, while the largest light neutrino 
mass ${m_\nu}_3$ is given by the $M_2$ effect.

\begin{figure}[t]
\begin{center}
\begin{minipage}{7.5cm}
\includegraphics[width=7cm]{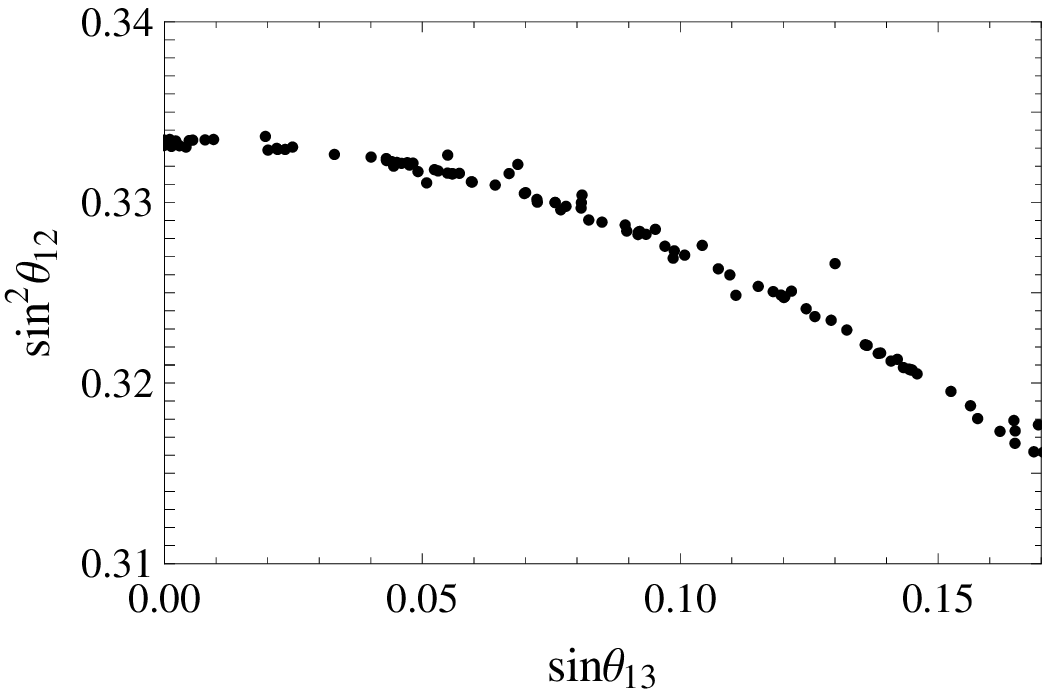}
\caption{Typical prediction for ($\theta_{12},\theta_{13}$) from the
cascade matrices.\bigskip}
\label{fig:1213}
\end{minipage}
\hspace*{5mm}
\begin{minipage}{7.5cm}
\includegraphics[width=7cm]{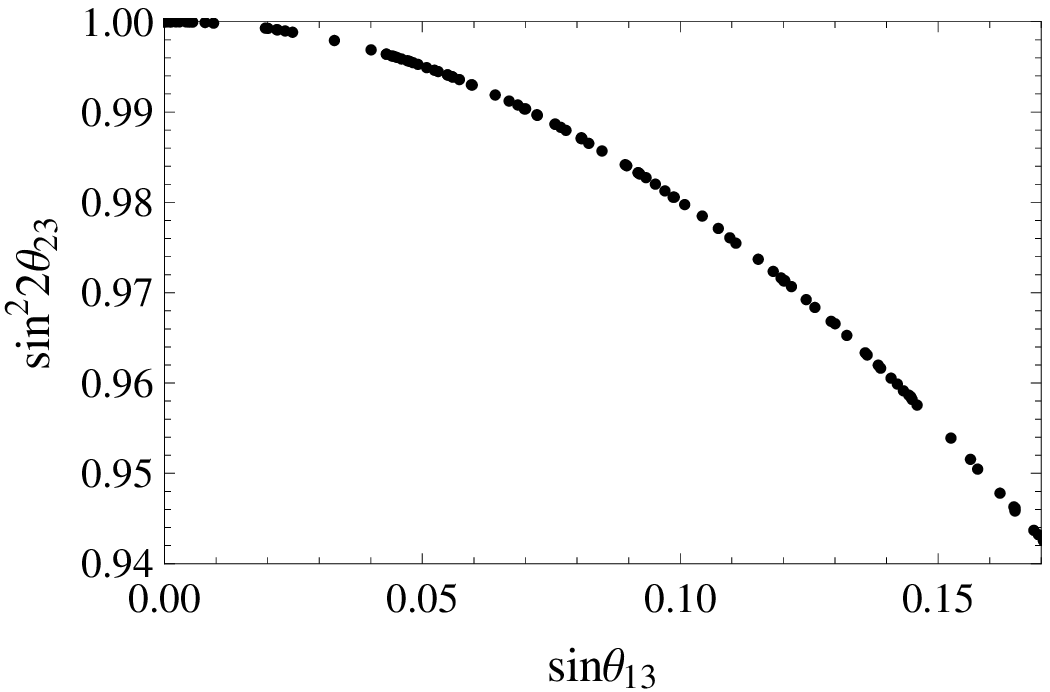}
\caption{Near maximal $\theta_{23}$ from the cascade matrices.\bigskip}
\label{fig:2313}
\end{minipage}
\end{center}
\end{figure}

It is seen from the above discussion that there are four combinations
of independent parameters while the five observed quantities
exist in the solar and atmospheric neutrino oscillations. Therefore
one parameter-independent relation among the observables is
found (the corrections from the charged-lepton sector will be
evaluated in the next section). That is explicitly written down in
the leading order of $r\equiv(\Delta m_{21}^2/\Delta m_{31}^2)^{1/2}$:
\begin{eqnarray}
  \frac{1}{9}\Big(\sin^2\theta_{23}-\frac{1}{2}\Big)
  -\frac{r}{4}\Big(\sin^2\theta_{12}-\frac{1}{3}\Big)
  -\frac{\sqrt{2}\,r}{27}\sin\theta_{13}
  \;=\; 0\,,
  \label{relation}
\end{eqnarray}
where we have taken the parameters as real valued. The 
relation \eqref{relation} is interpreted in two ways. First, the solar
neutrino angle $\theta_{12}$ has a correlation 
with $\theta_{13}$. Such behavior is given independently of the detail
of light neutrino mass spectrum. Figure~\ref{fig:1213} represents a
typical numerical calculation of $\theta_{12}$ and $\theta_{13}$.
It can be seen from the figure that $(\theta_{12},\theta_{13})$ is
concentrated near a thin curve implied by the 
relation \eqref{relation}. Secondly, the relation implies that the
atmospheric neutrino angle $\theta_{23}$ is near the maximal 
value $\pi/4$. We show in Fig.~\ref{fig:2313} the numerical evaluation
of the atmospheric neutrino angle. In both these figures, the
tri-bimaximal mixing is found to be realized around the central region
of parameter space.

\medskip

Finally we comment on other similar types of scenarios for neutrino
masses. First, for the asymmetric form of cascade, there is a solution
which makes the seesaw-induced mass matrix consistent 
with \eqref{MLTB}. In this solution, the 3-2 element in the Dirac 
mass $M_N$ is on the same order of the 3-3 element, like the so-called
lopsided matrix~\cite{lopside}. The inverted mass hierarchy of light
neutrinos is also viable for asymmetric cascades. Second, the
cascade form of the neutrino Dirac mass matrix is known to be preferred
from the viewpoint of parameter fine-tuning~\cite{natural}. The third 
scenario with a different principle is the sequential dominance
model~\cite{SD}. In this approach, the first law of $M_N$ is assumed
to have a vanishing element to realize the tri-bimaximal generation
mixing in the lepton sector. The choice of the vanishing element
depends on the order of right-handed neutrino masses. The sequential
dominance model has the parameter relations (\ref{cond}) in a
different basis, in other words, the mass hierarchy in $M_R$ is not
necessarily sequential. As for the (symmetric) cascade matrix
discussed in this paper, the hierarchical 
structure, $|\delta_i|\ll|\lambda_j|\ll1$, plays important roles for
realizing the tri-bimaximal mixing and neutrino mass
eigenvalues. Further we are motivated to explore the cascade 
form (\ref{cas}) for the extensions to the quark sector and also to
grand unified theory which connects the matrix forms of quarks and
leptons. The symmetric and hierarchical cascade also has a peculiar
dynamical origin, as will be discussed in a later 
section [e.g.\ see \eqref{3layers}].

\bigskip

\subsection{Charged-lepton sector}

As mentioned in the introduction, in this paper we explore the
possibility that the mass matrix of charged leptons also has the
cascade form. This is motivated, for example, by a high-energy
completion of the present framework with left-right gauge symmetry or
a more fundamental principle such as grand unification. In this
subsection, we study the corrections from the charged-lepton sector to
the lepton generation mixing angles.

Now the charged-lepton mass matrix takes the form:
\begin{eqnarray}
  M_E &=& \left(\begin{array}{ccc}
    \delta_1^e & \delta_2^e  & \delta_3^e \\
    \delta_2^e & \lambda_1^e & \lambda_2^e \\
    \delta_3^e & \lambda_2^e & 1 \\
   \end{array}\right)v_e,
\end{eqnarray}
where $\delta^e_1\sim\delta^e_2\sim\delta^e_3\sim{\cal O}(\delta^e)$
and $\lambda^e_1\sim\lambda^e_2\sim{\cal O}(\lambda^e)$. Unlike the
neutrino sector, the magnitudes of cascade hierarchy can be evaluated
from the experimentally observed values of charged-lepton masses and
given by
\begin{eqnarray}
  |\lambda^e| &\simeq& \frac{|m_\mu|}{|m_\tau|} \;\simeq\;
  6\times10^{-2},
  \label{lamE}  \\
  |\delta^e| &\simeq& \frac{|m_e|}{|m_\tau|} \;\simeq\; 3\times10^{-4}.
  \label{delE}
\end{eqnarray}
The generation mixing is expressed in terms of the cascade
hierarchy parameters, as shown in Table~\eqref{McasMwat}. Therefore
the corrections from the charged-lepton sector are found to be
generally small and the total lepton mixing angles are given at the
first order of perturbation
\begin{eqnarray}
  \sin^2\theta_{12} &=& \bigg|\frac{1}{\sqrt{3}}
  -\frac{2}{\sqrt{3}}\frac{{m_\nu}_1}{{m_\nu}_2}
  -\frac{1}{\sqrt{3}}\frac{m_e}{m_\mu}\bigg|^2, \\[1mm]
  \sin^2\theta_{23} &=& \bigg|\frac{-1}{\sqrt{2}} 
  +\frac{1}{\sqrt{2}}\frac{{m_\nu}_1(3{m_\nu}_3
    -{m_\nu}_2)}{{m_\nu}_3({m_\nu}_3-{m_\nu}_2)} 
  +\frac{\delta}{3\sqrt{2}\lambda}\,
  \frac{{m_\nu}_2}{{m_\nu}_3-{m_\nu}_2}
  -\frac{1}{\sqrt{2}}\frac{m_\mu}{m_\tau}\bigg|^2, 
  \label{23} \\[1mm]
  \sin^2\theta_{13} &=& \bigg|\frac{\delta}{\sqrt{2}\lambda}\,
  \frac{{m_\nu}_3-\frac{2}{3}{m_\nu}_2}{{m_\nu}_3-{m_\nu}_2}
  +\frac{\sqrt{2}\,{m_\nu}_1{m_\nu}_2}{{m_\nu}_3({m_\nu}_3-{m_\nu}_2)}
  +\frac{1}{\sqrt{2}}\frac{m_e}{m_\mu}\bigg|^2,
\end{eqnarray}From these expressions, 
one can see the effects of charged-lepton cascades. The solar neutrino
mixing is little (less than 1\%) affected and the tri-bimaximal solar
angle ($\sin^2\theta_{12}\simeq1/3$) still holds. As for the
atmospheric neutrino mixing, in the right-handed side of
Eq.~\eqref{23}, the charged-lepton effect often gives the dominant
correction ($\sim6\%$) to the tri-bimaximal atmospheric 
angle ($\sin^2\theta_{23}\simeq1/2$). Finally, the reactor neutrino
mixing sometimes receives a comparable effect relative to the neutrino
sector result. However its magnitude is of negligible 
order ($\lesssim1\%$) and the tri-bimaximal reactor 
angle ($\sin^2\theta_{13}\simeq0$) is not modified too much by the
charged-lepton sector. Since the hierarchy of the charged-lepton
cascade is expressed by the observables 
as \eqref{lamE} and \eqref{delE}, a parameter-independent relation
still holds including the charged-lepton correction:
\begin{eqnarray}
  \frac{1}{9}\Big(\sin^2\theta_{23}-\frac{1}{2}-\frac{m_\mu}{m_\tau}\Big)
  -\frac{r}{4}\Big(\sin^2\theta_{12}-\frac{1}{3}\Big)
  -\frac{\sqrt{2}\,r}{27}\sin\theta_{13}
  \;=\; 0\,,
\end{eqnarray}
in the first order approximation.

\bigskip

\section{Related phenomenology}
\label{pheno}

As we have shown, the cascade form of lepton mass matrices is well
fitted to the observed masses and mixing angles, and in particular,
yields the tri-bimaximal generation mixing from hierarchical mass
matrix structure. The non-trivial generation mixing in Yukawa matrices
generally provides rich flavor phenomenology other than fermion
masses. In this section, we investigate characteristic phenomenology 
induced by the cascade-form matrix: the lepton flavor violation in
supersymmetric extension of the theory, the baryon asymmetry of the
Universe via thermal leptogenesis, and the CP violation in neutrino
oscillations.

\subsection{Flavor violation}

First we estimate the branching ratios of flavor-violating rare decays
of charged leptons. In non-supersymmetric theory, the lepton flavor
violation (LFV) is suppressed and generally negligible because the
only source of low-energy LFV is the light neutrino masses and is very
small relative to the electroweak scale. On the other hand, the
supersymmetric (SUSY) theory generally predicts sizable magnitudes of
LFV amplitudes since additional sources of LFV come from mass
parameters of superparticles (scalar leptons). This type of
flavor-violating vertices are radiatively generated depending on the
form of lepton mass matrices. In the following, we estimate the
branching ratios of the rare decay 
processes $\,\ell_i\to\ell_j\gamma\,$ for the cascade lepton matrices.

We consider as a simple and conservative situation that soft
SUSY-breaking masses of scalar leptons are universal at some boundary
scale $\Lambda$. Then their off-diagonal matrix elements are generated
by radiative corrections from the Dirac Yukawa couplings of
neutrinos~\cite{LFV}. The one-loop renormalization group evolution
induces the left-handed scalar lepton masses which are approximately
given by
\begin{equation}
  ~~~ (m_\ell^2)_{ij} \;\sim\; \frac{1}{8\pi^2v^2}(3m_0^2+|a_0|^2)
  \sum_k (M_N^\dagger)_{ik}(M_N)_{kj}
  \ln\Big(\frac{|M_k|}{\Lambda}\Big), \quad\;
  ({\rm for} \;\;i\neq j)
  \label{ml2}
\end{equation} 
where $m_0$ and $a_0$ denote the universal SUSY-breaking mass and 
three-point coupling of scalar superpartners given at the boundary
scale $\Lambda$. The magnitude of these off-diagonal elements depend
on the form of Dirac neutrino mass matrix $M_N$ and the scale of
right-handed Majorana masses $M_i$. The expression \eqref{ml2} means
that the leading-order effects generally include large (i.e.\ the
third-generation) Yukawa couplings.

The branching ratio of the $\ell_i\to\ell_j\gamma$ process is given by
the loop diagrams including the vertex $(m_\ell^2)_{ij}$ in the mass 
insertion approximation. The result is roughly estimated as
\begin{equation}
  \text{Br}(\ell_i\to\ell_j\gamma) \;\sim\;
  \frac{3\alpha}{2\pi}\frac{|(m_\ell^2)_{ij}|^2M_W^4}{m_{\rm SUSY}^8}
  \tan^2\!\beta.
\end{equation}
Here $\alpha$ and $\tan\beta$ are the fine structure constant and the
ratio of two Higgs expectation values in supersymmetric SM,
respectively. In the denominator, $m_{\rm SUSY}$ denotes a typical
mass scale of superparticles circulating in the loops. In what
follows, we set $m_0=a_0=m_{\rm SUSY}$. Thus the branching ratios are
given in the table below:\smallskip
\begin{eqnarray}
{\renewcommand{\arraystretch}{1.4}%
\qquad\begin{array}{c||c|c} \hline
  & \text{Cascade} & \text{Waterfall} \\ \hline\hline
~\text{Br}(\mu\to e\gamma)~  &
 ~~C|\frac{m_1m_2}{m_3^2}|^2\big[\ln(\frac{|M_2|}{\Lambda})\big]^2~~ & 
 ~~C|\frac{m_1m_2}{m_3^2}|\,
 \big[\ln(\frac{|M_3|}{\Lambda})\big]^2~~ \\ \hline
 \text{Br}(\tau\rightarrow e\gamma) &
 C|\frac{m_1}{m_3}|^2\big[\ln(\frac{|M_3|}{\Lambda})\big]^2 &
 C|\frac{m_1}{m_3}|\,\big[\ln(\frac{|M_3|}{\Lambda})\big]^2 \\ \hline
 \text{Br}(\tau\to\mu\gamma) &
 C|\frac{m_2}{m_3}|^2\big[\ln(\frac{|M_3|}{\Lambda})\big]^2 &
 C|\frac{m_2}{m_3}|\,\big[\ln(\frac{|M_3|}{\Lambda})\big]^2 \\ \hline
\end{array}}
\end{eqnarray}
We have taken into account the charged-lepton corrections but these
are found to be negligible in the evaluation of LFV\@. The Dirac mass
eigenvalues $m_{1,2,3}$ are obtained by diagonalizing $M_N$. The
common factor $C$ is given by $C\simeq 10^{-5}B$ where $B$ is
determined model dependently by superparticle mass spectrum and Higgs
expectation values: $B\equiv(M_W/m_{\rm SUSY})^4\tan^2\!\beta$. For
comparison, we have listed in the table the results of waterfall-form
mass matrix. In particular, we obtain the following relations for the
cascade mass matrix:
\begin{eqnarray}
  \frac{\text{Br}(\mu\rightarrow e\gamma)}
       {\text{Br}(\tau\rightarrow\mu\gamma)}&\simeq&
  2|\lambda|^2\sin^2\theta_{13}
  \left[\frac{\ln(|M_2|/\Lambda)}{\ln(|M_3|/\Lambda)}\right]^2,
  \label{muetaumu} \\
  \frac{\text{Br}(\tau\rightarrow e\gamma)}
       {\text{Br}(\tau\rightarrow\mu\gamma)}&\simeq&
  2\sin^2\theta_{13},
\end{eqnarray}
where the reactor angle $\theta_{13}$ is given by the neutrino sector
contribution.\footnote{The sequential dominance model, mentioned in
Section~\ref{cas_lep}, presents somewhat different phenomenological
predictions. For instance, 
$\text{Br}(\mu\rightarrow e\gamma)/\text{Br}(\tau\rightarrow\mu\gamma)
\propto\sin^2(2\sqrt{2}\theta_{13})$ when $M_3$ is the heaviest
right-handed neutrino mass. Further, since the sequential dominance
model assumes that the charged-lepton generation mixing is similar to
that of quarks in the sense that the mixing is dominated by the 1-2
mixing, the reactor angle $\theta_{13}$ is determined by the
correction from the charged-lepton sector. These facts are compared 
with the predictions from the cascade mass matrix: the ratio given 
in (\ref{muetaumu}) depends on the hierarchy of neutrino mass cascade,
and also the charged-lepton correction to $\theta_{13}$ is found to be
small.}
\begin{figure}[t]
\begin{center}
\includegraphics[width=11cm]{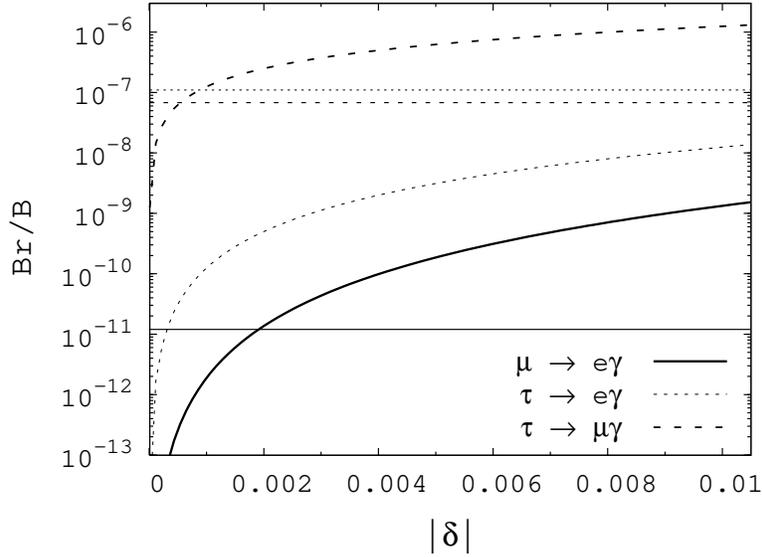}
\caption{Typical predictions for lepton flavor violation in the
cascade model. The solid, dotted, and dashed lines denote the
branching ratios (over the common $B$ factor defined in the 
text) of $\mu\to e\gamma$, $\tau\to e\gamma$, and $\tau\to\mu\gamma$,
respectively. The cascade hierarchy is fixed $|\delta|=|\lambda|^2$ in
the figure. The corresponding horizontal lines mean the current
experimental upper bounds.\bigskip}
\label{fig:LFV}
\end{center}
\end{figure}
The dominant contribution to $\text{Br}(\mu\to e\gamma)$ comes from
the second-generation effect in the cascade model, while all the
other branching ratios depend on $M_3$ due to the large Yukawa
coupling of the third generation. It is found from the above table
that, for fixed mass eigenvalues, all the LFV processes in the cascade
model are more suppressed than the waterfall model. The suppression is
enough, even when $\tan\beta$ is large or the superparticle mass 
scale $m_{\rm SUSY}$ is around the electroweak scale. For example,
if $|\delta|=|\lambda|^2=10^{-4}$, typical Majorana masses given 
in \eqref{M1}-\eqref{M3} read
\begin{equation}
  |M_1| \,\sim\, 10^8 \mbox{ GeV}, \qquad
  |M_2| \,\sim\, 10^{11} \mbox{ GeV}, \qquad
  |M_3| \,\sim\, 10^{16} \mbox{ GeV}, 
\end{equation}
and the branching ratios then become
\begin{gather}
  \text{Br}(\mu\to e\gamma) \,\sim\, 10^{-15}B, \qquad
  \text{Br}(\tau\to e\gamma) \,\sim\, 10^{-12}B, \qquad
  \text{Br}(\tau\to\mu\gamma) \,\sim\, 10^{-8}\,B, \\[2mm]
  B \,=\, \Big(\frac{M_W}{m_{\rm SUSY}}\Big)^4\tan^2\!\beta.
  \nonumber
\end{gather}
These results are compared with the current experimental upper bounds
at the 90\% confidence level~\cite{PDG}:
$\text{Br}(\mu\to e\gamma)<1.2\times10^{-11}$,
$\,\text{Br}(\tau\to e\gamma)<1.1\times10^{-7}$, and
$\text{Br}(\tau\to\mu\gamma)<6.8\times10^{-8}$.
\begin{figure}[t]
\begin{center}
\includegraphics[width=11cm]{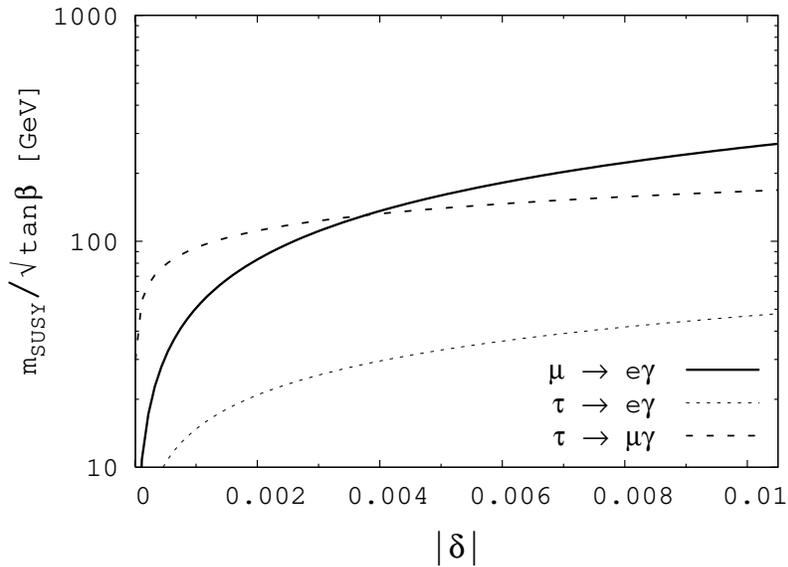}
\caption{Typical lower bounds of superparticle mass scale which come
from the LFV processes, $\mu\to e\gamma$ (solid 
line), $\tau\to e\gamma$ (dotted line), and $\tau\to\mu\gamma$ (dashed
line). The cascade hierarchy is fixed as $|\delta|=|\lambda|^2$ in the
figure.\bigskip}
\label{fig:msusy}
\end{center}
\end{figure}
The first two predictions are far below the experimental limits. On
the other hand, the $\tau\to\mu\gamma$ decay is marginal to the
present bound and would be observed in future LFV searches with
relatively light superparticle spectrum. The branching ratios
increase as the cascade factors $\delta$ and $\lambda$, and larger
values of these factors lead to observable effects as shown in
Fig.~\ref{fig:LFV}. This fact in turn constrains the mass scale of
superparticles. In Fig.~\ref{fig:msusy}, we show the lower bound of
SUSY scale $m_{\rm SUSY}$ for a typical hierarchy in the cascade
neutrino matrix ($|\delta|=|\lambda|^2$)\@. The figure shows 
that, for $|\delta|\gtrsim3\times10^{-3}$ ($\lesssim3\times10^{-3}$),
the experimental limit from the $\mu\to e\gamma$ ($\tau\to\mu\gamma$)
decay imposes the most severe constraint on the SUSY-breaking scale,
while the $\tau\to e\gamma$ decay rate is too small to be
detected. For larger hierarchy of the 
cascade, $|\delta|=|\lambda|^n$ ($n>2$), the $\mu\to e\gamma$ decay is
more suppressed but the $\tau\to\mu\gamma$ process is not. Therefore
the lower bound on $m_{\rm SUSY}$ is weakened and becomes
insignificant for larger values of $|\lambda|\gtrsim5\times10^{-2}$.

\bigskip

\subsection{CP violation}

Next let us study CP-violating phenomenology, in particular, examine
whether the thermal leptogenesis~\cite{FY} works in the cascade
model. The CP-asymmetry parameter in the decay of the right-handed
neutrino $R_i$ is defined as
\begin{equation}
  \varepsilon_i \;=\;
  \frac{\sum_j \Gamma(R_i\rightarrow L_j H)-
  \sum_j \Gamma(R_i\rightarrow L_j^c H^\dagger)}{
  \sum_j \Gamma(R_i\rightarrow L_j H)+
  \sum_j \Gamma(R_i\rightarrow L_j^c H^\dagger)}\,.
\end{equation}
As seen in the previous section, the cascade model has hierarchical
mass eigenvalues of right-handed neutrinos. In this case,
neglecting thermal corrections, an approximate formula 
for $\varepsilon_1$ at low temperature (but reasonably accurate even
at higher temperatures) is given by~\cite{LPG}
\begin{equation}
  \varepsilon_1 \;=\; \frac{1}{8\pi}
  \sum_{j\neq1}\frac{{\rm Im}(A_{j1})^2}{|A_{11}|}F(r_j)
  \label{ep}
\end{equation}
in the basis that the right-handed Majorana mass matrix is
diagonalized (with real positive eigenvalues). The mass ratios of
right-handed neutrinos are denoted 
by $r_j\equiv |M_j/M_1|^2$. The hermite matrix $A$ is defined 
as $A\equiv(DM_NM_N^\dagger D^\dagger)/v^2$ where $D$ is the diagonal
phase matrix which makes the eigenvalues $M_i$ real and positive. The
loop function $F$ is determined by evaluating the Feynman diagrams for 
the $R_1$ decay; 
\begin{equation}
  F(x) \;=\; \left\{\begin{array}{ll}
  \sqrt{x}\,\bigg[\,\dfrac{2-x}{1-x}-
  (1+x)\ln\Big(1+\dfrac{1}{x}\Big)\bigg] &\qquad \text{(SM)} \\[4mm]
  \sqrt{x}\,\bigg[\,\dfrac{2}{1-x}-
  \ln\Big(1+\dfrac{1}{x}\Big)\bigg] &\qquad \text{(SUSY SM)}
  \end{array}\right.
\end{equation}
Note that the loop function factor $F(r_j)$ behaves 
as $1/r_j^{1/2}$ for large mass hierarchy, i.e.\ $r_j\gg1$. The
relevant quantities for $\varepsilon_1$ are listed in the table
below:\smallskip
\begin{eqnarray}
{\renewcommand{\arraystretch}{1.3}%
\quad\begin{array}{c||c|c|c|c|c} \hline
  & ~~A_{11}~~ & ~~|A_{12}|~~ & ~|A_{13}|~ & 
  ~M_1/M_3~ & ~M_2/M_3~ \\ \hline\hline
\text{Cascade}~ & 3|\delta|^2 & \,|\delta|^2 & |\delta| & 
{\cal O}(\delta^2) & {\cal O}(\lambda^2) \\  \hline
\text{Waterfall}~ & \,|\delta|^2 & |\delta\lambda| & |\delta| &
{\cal O}(\delta^3\lambda) & {\cal O}(\delta\lambda^3)  \\  \hline
\end{array}}
\end{eqnarray}
It is found from the definition of $\varepsilon_1$ (or the matrix $A$)
that the charged-lepton effect, i.e.\ the left-handed field rotation,
does not change the CP asymmetry and need not be included. The
generation mixing of Dirac neutrinos is set to be of similar order
between the two types of matrices and so the hierarchy flow in the 
right-handed Majorana mass matrix becomes more rapid in the waterfall
model. This fact leads to the result that the cascade model generally
predicts larger cosmological CP asymmetry than the waterfall
model. However notice that $A_{12}$ in the cascade model is not a
naive expectation ${\cal O}(\delta\lambda)$ but a suppressed 
value ${\cal O}(\delta^2)$. This is because of a cancellation caused 
by the relative sign, $\lambda_1=-\lambda_2$, which is suggested by the
current neutrino experimental data [see the third term in the mass
matrix \eqref{MLTB}]. Consequently, the effect of the second
generation often becomes sub-leading in the cascade model, as will be
seen in the following. On the other hand, it is found from the above
table that in the waterfall model the second-generation effect is
dominant. The hierarchy factor dependence of the CP asymmetry is
roughly estimated by dropping numerical factors 
as $\varepsilon_1\sim(A_{13}^2/A_{11})(M_1/M_3)
\sim{\cal O}(\delta^2)$ in the cascade model 
and $\varepsilon_1\sim(A_{12}^2/A_{11})(M_1/M_2)
\sim{\cal O}(\delta^2)$ in the waterfall model. Therefore the cascade
hierarchy is found to generally induce similar or sometimes larger
baryon asymmetry compared with the Froggatt-Nielsen like hierarchy.

From the general formula \eqref{ep}, we obtain the asymmetry parameter
for the SM with the cascade mass matrix;
\begin{equation}
  \varepsilon_1 \;\simeq\;
  \frac{-1}{16\pi|\delta|^2}\bigg[\,
    |\delta|^4\sin(\theta_2-\theta_1)\bigg|\frac{M_1}{M_2}\bigg|+
    {\rm Im}\big[\delta^2e^{i(\theta_3-\theta_1)}\big]
    \bigg|\frac{M_1}{M_3}\bigg|\,\bigg],
  \label{lpgep}
\end{equation}
where $\theta_i=\arg(M_i)$. In supersymmetric extensions, the result
becomes twice that of the SM because the loop function $F$ differs by
a factor 2 when the right-handed neutrino masses are 
hierarchical; $|M_{2,3}|\gg|M_1|$. Moreover the decay of the
superpartner of $R_1$ also generates roughly the same size of
asymmetry as \eqref{lpgep} due to the presence of supersymmetry. Given
the mass hierarchy of right-handed 
neutrinos \eqref{M1}-\eqref{M3} (with the best-fit values of neutrino
oscillation parameters), the ratio of the first and second terms 
in \eqref{lpgep} is found to be 
$\sim50\times|\delta^2|/|\lambda^2|$, and hence the second term
is dominant unless $M_3$ is huge or $\delta$ has a particular
value so that $\arg(\delta^2)=\theta_1-\theta_3$. We here define the
resultant CP asymmetry $\eta_{\rm CP}^{}$ as the ratio of the lepton
asymmetry and the photon number density $n_\gamma$. This is
parameterized as
\begin{equation}
  \eta_{\rm CP}^{} \;=\; \frac{135\,\zeta(3)}{4\pi^4}
  \frac{\kappa s}{g_*}\frac{\varepsilon_1}{n_\gamma},
  \label{etaCP}
\end{equation} 
where $s$ is the entropy density and $g_*$ is the effective number of
degrees of freedom in thermal equilibrium; $s=7.04\,n_\gamma$ in the
present epoch and $g_*=106.75$ ($228.75$) for the SM (for the minimal
SUSY SM)\@. The numerical factor in \eqref{etaCP} denotes the
equilibrium $R_1$ number density relative to the entropy density. As
mentioned above, in supersymmetric theory the scalar neutrino decay
roughly doubles the result \eqref{etaCP}.

\begin{figure}[t]
\begin{center}
\includegraphics[width=11cm]{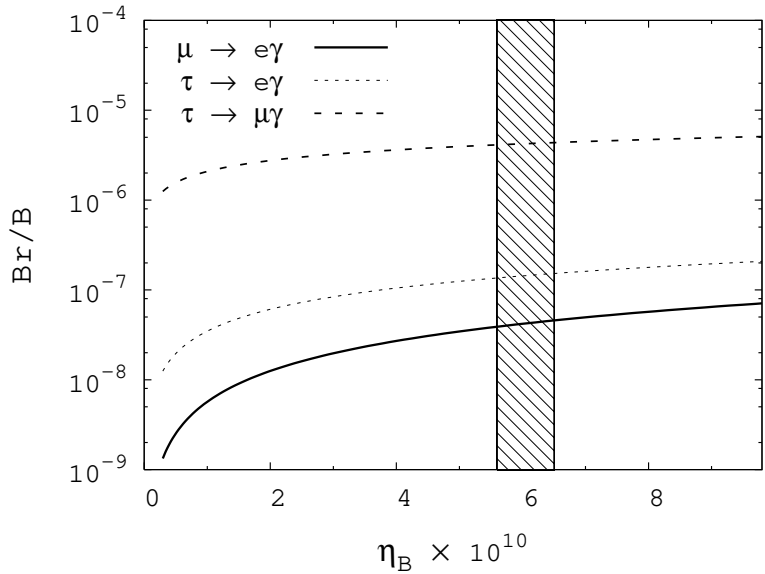}
\caption{Phenomenology of the cascade lepton matrices: the flavor
violation and the baryon asymmetry. The cascade hierarchy parameters
are around $|\delta|=|\lambda|^2\sim{\cal O}(10^{-2})$ in the
figure. A relative complex phase between the second and
third-generation effects in \eqref{lpgep} is set to be
constructive. The vertical shadow band is allowed by the cosmological
observations at the 95\% CL.\bigskip}
\label{fig:LPGLFV}
\end{center}
\end{figure}

The efficiency factor $\kappa$ is obtained by numerically solving the
Boltzmann equations and is a function of two parameters: the heavy
mode mass $M_1$ and the effective light neutrino 
mass $m_{\rm eff}\equiv|(M_N^\dagger M_N)_{11}/M_1|$. In particular,
the efficiency is known to depend only 
on $m_{\rm eff}$ when $|M_1|\ll10^{14}$ GeV, which is realized in the
cascade model [see \eqref{M1}]\@. That leads to an approximate
formula~\cite{LPG2}:
\begin{equation}
  \kappa^{-1} \;\simeq\; \frac{3.3\times10^{-3}\,\text{eV}}{m_{\rm eff}}
  +\bigg(\frac{m_{\rm eff}}{0.55\times10^{-3}\,\text{eV}}\bigg)^{1.16},
  \label{kappa}
\end{equation}
with vanishing initial $R_1$ population. We have shown in the previous
section that the cascade neutrino mass matrix leads 
to $m_{\rm eff}=|\hat m_2|\simeq\sqrt{\Delta m^2_{21}}\sim10^{-2}
\,\text{eV}$, and therefore the second term in \eqref{kappa} becomes
dominant. The baryon number asymmetry $\eta_B^{}$ is transfered via
spharelon interactions 
as $\eta_B^{}=-\frac{28}{79}\eta_{\rm CP}^{}$ in the SM 
and $\eta_B^{}=-\frac{8}{23}\eta_{\rm CP}^{}$ in the minimal SUSY
SM\@. Combining the above result and the mass parameters calculated
previously, we obtain the baryon asymmetry of the Universe in the
cascade model [taking account only of the leading (second) term 
in \eqref{lpgep}]:
\begin{equation}
  \eta_B^{} \;\leq\; 8.4\times10^{-6}\,|\delta|^2\sin\theta_B,
  \label{etaB}
\end{equation}
where $\theta_B=2\arg(\delta)+\theta_3-\theta_1$. An almost similar
size of $\eta_B^{}$ is obtained for the SUSY SM, considering the fact
that the washout effect is two times stronger because of the additional
decay channel to superpartners. The current experimental data at 95\%
confidence level shows that $\eta_B^{}=(4.7-6.5)\times10^{-10}$ from
the big bang nucleosynthesis 
result~\cite{PDG} and $\eta_B^{}=(5.6-6.5)\times10^{-10}$ from 
the WMAP 3-year mean result in the standard $\Lambda$ cold dark matter
scenario~\cite{WMAP}. Then it is found that the magnitude of neutrino
cascade hierarchy,
\begin{equation}
  |\delta| \;\gtrsim\; (7.5-8.8)\times10^{-3},
  \label{dbound}
\end{equation}
is consistent with the baryon asymmetry of the Universe 
with ${\cal O}(1)$ complex phases of mass parameters. In fact, for a
larger value of the cascade hierarchy, the first term 
in \eqref{lpgep} becomes effective and must be taken into account in
the analysis. Consequently the above bound of $|\delta|$ is modified
by an ${\cal O}(1)$ factor. In Fig.~\ref{fig:LPGLFV}, we plot the full 
numerical evaluation of phenomenological consequences of the cascade
lepton matrices.
In this figure, the cascade hierarchy parameters are 
around $|\delta|=|\lambda|^2\sim{\cal O}(10^{-2})$. It is found that,
for this type of hierarchy, the $\mu\to e\gamma$ rare decay process
implies a lower bound on the superparticle mass scale to be larger than
about $450\sqrt{\tan\beta}$ GeV, which is a bit heavier than scalar
leptons in typical minimal supergravity scenarios. Note that, for
a larger hierarchy of the cascade, $|\delta|=|\lambda|^n$ ($n>2$), the
only modification is more suppression of the first term 
in \eqref{lpgep} and the above result is affected little.

The asymmetry $\eta_B^{}$ becomes tiny in the limit of a 
vanishing ${m_\nu}_1$ or equivalently a huge $M_3$. In other words, an 
upper bound on $M_3$, for example $|M_3|<M_{\rm GUT}$, leads to a
restricted prediction of the baryon asymmetry. The maximal value of
asymmetry shown in \eqref{etaB} is obtained in the case of a smaller
solar angle and a smaller $\Delta m_{21}^2$ within the range of
experimental bounds. The corresponding lower 
bound \eqref{dbound} means that the Yukawa hierarchy in the neutrino
sector is a bit smaller than in the charged-lepton 
sector ($|\delta|>|\delta^e|\sim m_e/m_\tau$). Such a hierarchy factor
may be reduced, for example, by taking a larger value 
of $|{m_\nu}_1/{m_\nu}_2|$ with suitable complex phases. Another
reasonable possibility is to consider different initial population 
of the right-handed neutrino $R_1$ at high temperature. More abundance
of initial $R_1$ makes the efficiency factor larger. For example, if
we choose the cascade factor $\delta$ as the same order 
of $\delta^e$, then $|M_1|\sim10^8$ GeV and the efficiency 
factor $\kappa$ is enhanced by 1-3 orders of magnitude, depending on
the initial $R_1$ abundance. This behavior makes $\eta_B^{}$ enhanced
and reduces $\delta$ by 1 or 2 orders of magnitude, which is
consistent with $|\delta|\lesssim|\delta^e|$.

\medskip

Another important CP-violating phenomenon is the Dirac-type CP
violation in neutrino oscillations, which could be observed in future
long baseline experiments. The effect of the Dirac CP phase is
expressed in terms of the quantity $J_{\rm CP}$ which is invariant
under the rephasing and relabeling of fermion fields~\cite{J}. From
the analysis of the generation mixing matrix in the previous section,
we find that $J_{\rm CP}$ from the cascade lepton mass matrices is
given by
\begin{eqnarray}
  J_{\rm CP} \;=\; \frac{1}{6}\,\text{Im}
  \left[\,\frac{\delta}{\lambda}\cdot
    \frac{{m_\nu}_3-\frac{2}{3}{m_\nu}_2}{{m_\nu}_3-{m_\nu}_2}
    +\frac{2{m_\nu}_1{m_\nu}_2}{{m_\nu}_3({m_\nu}_3-{m_\nu}_2)}
    +\frac{m_e}{m_\mu}\,\right],
  \label{Jcp}
\end{eqnarray}
in the leading approximation. The maximum value of $J_{\rm CP}$ is
related to the LFV branching ratios as $\text{Br}(\tau\to e\gamma)/
\text{Br}(\tau\to\mu\gamma)=(6J_{\rm CP}^{\rm max})^2$. Further, from
the cosmological analysis, the contribution of the first term 
in \eqref{Jcp} is found to be dominant. Thus the CP violation in
neutrino oscillations is approximately described by the phase of
cascade hierarchy factor; $\arg(\delta/\lambda)$. As shown
hereinbefore, the cascade form of lepton mass matrices leads to
characteristic and correlated behaviors for flavor physics and
cosmology. That deserves to be investigated in more detail and
examined in future particle experiments.

\bigskip

\section{Illustrative toy models}
\label{dynamics}

In this section we show that the cascade form of mass (Yukawa) matrix
has possible dynamical origins in high-energy regime. The cascade
contains two step hierarchies of the orders 
of $\delta$ and $\lambda$. The former factor is concerned with the
first generation and the latter with the second one. Further, as
argued above, the neutrino experimental data would suggest that the
coefficients of effective mass (Yukawa) operators are correlated to
each other. These non-trivial properties imply some non-trivial
implements introduced in fundamental theory beyond the standard
model.

\medskip

\subsection{Flavor symmetry}

The first example is to introduce an abelian flavor symmetry. The
standard model and its extensions contain three-generation left and
right-handed fermions, $L_i$ and $R_i$ ($i=1,2,3$), and the Higgs
field which has a non-vanishing vacuum expectation value $v$. In
addition to these fundamental fields, here three gauge-singlet 
scalars $\phi_j$ ($j=1,2,3$) are also included. We write down the
model in a supersymmetric way using $L$, $R$, $\phi$ as corresponding
superfields, but a non-SUSY theory is easy to construct with an
additional symmetry which reflects the holomorphicity of
superpotential terms. The quantum number assignment of $U(1)$ flavor
symmetry is determined in the following way:\smallskip
\begin{eqnarray}
{\renewcommand{\arraystretch}{1.3}%
\begin{array}{c||c|c|c|c|c|c|c|c|c} \hline 
& L_1 & ~L_2~ & ~L_3~ & R_1 & ~R_2~ & ~R_3~ & 
\phi_1 & ~\phi_2~ & ~\phi_3~ \\
\hline\hline
~U(1)~ & 2m+1 & 1 & 0 & 2m+1 & 1 & 0 & -2m-3 & -2 & -1 \\ \hline
\end{array}}
\label{U1}
\end{eqnarray}
where $m$ is an arbitrary positive integer. We have taken the 
matter $U(1)$ charges as symmetric ones ($Q_{L_i}=Q_{R_i}$) and the
third-generation fields have zero charges ($Q_{L_3,R_3}=0$). The
latter fact just defines the overall scale of induced mass terms,
which scale can be easily reduced by a universal shifting of all
charges so that $Q_{L_3},Q_{R_3}>0$.

The effective mass terms come from the operators which are consistent
with the flavor symmetry and are generally higher dimensional
suppressed by the cutoff scale $\Lambda$, at which the operators are
effectively generated by high-scale dynamics. The induced Dirac
mass operators in the superpotential $\,W=R_i(M_D)_{ij}L_j$ are now
given by
\begin{eqnarray}
  M_D &=& \left(\begin{array}{ccc}\!
    \dfrac{\phi_1\phi_2^{m-1}\phi_3}{\Lambda^{m+1}} &
    \dfrac{\phi_2^{m+1}}{\Lambda^{m+1}} &
    \dfrac{\phi_2^m\phi_3}{\Lambda^{m+1}}\! \\[2mm]
    \dfrac{\phi_2^{m+1}}{\Lambda^{m+1}} &
    \dfrac{\phi_2}{\Lambda} & \dfrac{\phi_3}{\Lambda} \\[2mm]
    \dfrac{\phi_2^m\phi_3}{\Lambda^{m+1}} & \dfrac{\phi_3}{\Lambda} & 1
  \end{array}\right)v.
\end{eqnarray}
It would be expected that these scalar fields develop the same
magnitude of expectation 
values $\langle\phi_1\rangle\simeq\langle\phi_2\rangle
\simeq\langle\phi_3\rangle\equiv\lambda\Lambda$, e.g.\ governed by a
single sector dynamics, and as a result the mass matrix becomes
\begin{eqnarray}
  M_D \;\simeq\; \left(\begin{array}{ccc}
    \lambda^{m+1} & \lambda^{m+1} & \lambda^{m+1} \\
    \lambda^{m+1} & \lambda & \lambda \\
    \lambda^{m+1} & \lambda & 1
  \end{array}\right)v.
\end{eqnarray}
This is the cascade-form matrix with the 
hierarchy $\delta\simeq\lambda^{m+1}$ ($m$ is an arbitrary positive
integer). In fact, the quantum number assignment \eqref{U1} is shown
to be unique, up to an overall rescaling, in the case that one flavor
symmetry and three gauge-singlet fields generate the cascade.

For the neutrino sector, the Majorana mass matrix of right-handed
neutrinos was taken to be flavor diagonal in the previous
analysis. That is realized, e.g.\ by introducing several scalars which
transform non-trivially under additional symmetry (the lepton number
or some discrete symmetry). It is however noted that, as we mentioned
before, the right-handed Majorana mass matrix can also be of the
cascade form, which is derived in a similar way to the above.

\bigskip

\subsection{Extra dimensions}

The second example is an extension of the SM involving the extra
spacetime beyond our four dimensions and the non-abelian discrete
flavor symmetry as its heritage.

An interesting key to realize the cascade form is the observation that
the cascade is split into three layers (see also Fig.\ref{fig:matrices}):
\begin{eqnarray}
  M_{\rm cas} &\propto&
  \left(\begin{array}{ccc}
    \delta & \delta & \delta \\
    \delta & \delta & \delta \\
    \delta & \delta & \delta
  \end{array}\right) +
  \left(\begin{array}{ccc}
    \;\,& & \\
    & \lambda & \lambda\! \\
    & \lambda & \lambda\!
  \end{array}\right) +
  \left(\begin{array}{ccc}
    \;\, & \;\, & \\
    & & \\
    & & 1\!
  \end{array}\right).
  \label{3layers}
\end{eqnarray}
The first and second terms indicate the existence of non-abelian
flavor symmetry which leads to generation-correlated values of matrix
elements. The first two terms also contain the suppression factors
relative to the last term. The dynamical origin of suppression is here
traced to the dilution of existence probabilities in the extra spatial
dimensions.

Let us consider a six-dimensional theory on the flat gravitational
background. The extra two-dimensional space is compactified on the
torus $T^2$ with the radii $R_5$ and $R_6$. In addition, the theory is
assumed to have the $Z_3$ invariance which acts as 
the $2\pi/3$ rotation on $T^2$. That implies the torus is the 
diamond ($R_5=R_6\equiv R$) with an interior angle $2\pi/3$. The torus
is further divided by $Z_3$ and results in the orbifold $T^2/Z_3$. The
orbifold has three $Z_3$ fixed 
points: $P_1=(0,2\pi R/\sqrt{3})$, $P_2=
(\pi R,\pi R/\sqrt{3})$, and $P_3=(0,0)$. The assertion of the
equivalence of three fixed points may lead to the existence of
permutation $S_3$ flavor symmetry in the low-energy effective theory
of this setup~\cite{ExDS3}.

We briefly show a schematic picture of field configuration in the
extra dimensions. The three-generation left and right-handed fermions
are assumed to be generation-separately localized on the three fixed
points $P_{1,2,3}$ of the orbifold. As for the bosonic sector of the
theory, we show a simple example that the electroweak Higgs 
field $H$ comes from a six-dimensional scalar, and further three types
of gauge-singlet scalars are arranged, $\phi_1$ in the 
bulk, $\phi_2$ on a line, and $\phi_3$ on a fixed point: the latter
fixed point means $P_3$ corresponding to the third generation and the
line connects two fixed points $P_2$ and $P_3$ on which the second and
third-generation fermions reside.

The six-dimensional scalar $\phi_1$ couples to all the
three-generation fermions. If the effective theory has independent
flavor symmetry for left and right-handed fermions,\footnote{For
details of permutation flavor symmetry, see for example
Ref.~\cite{S3}.} the operators involving $\phi_1$ induce the lowest
layer of the cascade [the first term in the cascade 
matrix \eqref{3layers}] with the universal coefficient. The 
scalar $\phi_2$, which extends into the fifth dimension, is separate
from the first-generation fields due to the locality in the extra
dimensions. That results in producing the middle layer of the
cascade. The coefficients of $\phi_2$ operators may be controlled by
a subgroup of flavor symmetry. Similarly, the four-dimensional 
scalar $\phi_3$ couples to the third generation on the same fixed
point, and hence induces the third term in the cascade 
matrix \eqref{3layers}. As for the hierarchy (the relative heights of
the cascade layers), it has an interesting dynamical origin in the
present framework: it is determined by the volume of extra-dimensional
space. That is, the scalar fields $\phi_{1,2,3}$, which generate
effective Yukawa operators, have different dimensionality and then
provide different volume suppression factors for mass matrix
elements. In the above example, the relative hierarchies are given 
by $\delta\simeq1/\Lambda R$ and $\lambda\simeq1/\sqrt{\Lambda R}$
where $\Lambda$ is the cutoff scale of the theory ($\gg1/R$). It is
possible to have a larger hierarchy, $\delta\sim\lambda^n$ ($n>2$),
if the scalar $\phi_1$ extends to more higher-dimensional spacetime.

Finally we comment on the Majorana mass term for right-handed
neutrinos. It can be obtained in a similar way to the above by
introducing extra scalar fields with different dimensionality. If
these scalars have the lepton number one, the mass hierarchy in the
right-handed Majorana matrix is the square of that in the Dirac
one. This fact realizes in a dynamical way the 
result \eqref{M1}-\eqref{M3} with the central values of experimental
data.

The higher-dimensional framework has a variety of possible field
configurations in different classes of extra-dimensional space, each
of which has an individual low-energy prediction. Other types of
configurations are then constructed to realize the cascade-form
matrix: a six-dimensional theory compactified on $T^2/Z_3$ with
three-generation fermions being localized on three different
lines, a seven-dimensional theory compactified on a torus or
octahedron with three generations extending to different directions of
extra three spatial dimensions, etc. The scheme given in this
subsection is just for illustration and an explicit construction of
a realistic complete theory is left for future study.

\bigskip

\section{Summary}
\label{summary}

In this paper we have investigated the phenomenology of cascade mass
matrices in the neutrino and charged-lepton sectors. Implementing the
seesaw mechanism, an approximate tri-bimaximal generation mixing is
found to be induced from hierarchical lepton mass matrices. The
flat-cascade lepton matrices, which are well fitted to the
experimental data, imply one parameter-independent relation among the
observables, the generation mixing angles and mass eigenvalues. The
relation means the correlated values of mixing angles, that behavior
will be tested in future neutrino experiments.

We have discussed several phenomenological aspects of the neutrino and
charged-lepton mass matrices in the flat-cascade form. The first is
the flavor-violating rare decay of charged leptons in supersymmetric 
standard models. While the branching ratios are suppressed due to
small flavor mixing with fixed mass eigenvalues, several decay modes
give observable effects and in turn impose the lower bound on the
supersymmetry-breaking scale. The second is the CP-violating phenomena
in neutrino oscillations and cosmology. The latter means the baryon
asymmetry of the Universe via the leptogenesis produced in the cascade
model. The predictions of these quantities are found to be correlated
and make the cascade-form mass matrix testable in near future.

We have also illustrated several dynamical frameworks for realizing the
cascade-form matrix. The dynamics involves the existence of flavor
symmetry and/or extra spatial dimensions. Along the lines presented
here, the construction of realistic model including the quark sector
and the investigation of induced phenomenology will be the next
important tasks to probe the existence of cascades in nature.

\bigskip\bigskip

\subsubsection*{Acknowledgments}

N.H.\ and M.T.\ have been supported in part by scientific grants
from the Ministry of Education, Science, Sports, and Culture of 
Japan (No.~16540258 and No.~17740146) and (No.~17540243 and 
No.~19034002). K.Y.\ is supported by the grant-in-aid for scientific 
research on the priority area (\#441) ``Progress in elementary
particle physics of the 21st century through discoveries of Higgs
boson and supersymmetry'' (No. 16081209) and by a scientific grant
from Monbusho (No.~17740150). The work of R.T.\ has been supported by
the Japan Society of Promotion of Science.

\newpage

\end{document}